\documentclass[letterpaper]{report}
\usepackage[utf8]{inputenc}
\usepackage[T1]{fontenc}
\usepackage{arxivWrapper}
\usepackage{amsmath,amssymb,array}
\usepackage{booktabs}

\usepackage{bm}

\renewcommand{\thechapter}{\Roman{chapter}}     
\makeatletter                                   
\renewcommand{\p@section}{\thechapter.}         
\makeatother                                    

\newcommand{\tree}{\mathcal{T}}

\begin{document}
\sectionhead{}
\volume{}
\volnumber{}
\year{}
\month{}

\begin{article}
  \title{Structured Bayesian Regression Tree Models for Estimating Distributed Lag Effects: The R Package {dlmtree}}
\author{by Seongwon Im, Ander Wilson, and Daniel Mork}

\maketitle

\abstract{
When examining the relationship between an exposure and an outcome, there is often a time lag between exposure and the observed effect on the outcome. A common statistical approach for estimating the relationship between the outcome and lagged measurements of exposure is a distributed lag model (DLM). Because repeated measurements are often autocorrelated, the lagged effects are typically constrained to vary smoothly over time. A recent statistical development on the smoothing constraint is a tree structured DLM framework. We present an R package dlmtree, available on CRAN, that integrates tree structured DLM and extensions into a comprehensive software package with user-friendly implementation. A conceptual background on tree structured DLMs and demonstration of the fitting process of each model using simulated data are provided. We also demonstrate inference and interpretation using the fitted models, including summary and visualization. Additionally, a built-in shiny app for heterogeneity analysis is included.
}

\section{Introduction}

In many fields, there is interest in estimating the lagged relationship between an exposure (or treatment) and an outcome. In such cases, the length of the lag or how the exposure effect varies is often unknown. The lagged relationship can take either of two forms. The first form is the association between the exposure at one time point and an outcome distributed across several subsequent times. For example, the impact of advertisement persists beyond that single time point of the investment \citep{koyck_1954_econDLM, palda_1965_advertise}
or exposure to an environmental pollutant affects mortality on the same day and each of the following days \citep{schwartz_distributed_2000}. The second form is an exposure assessed longitudinally and an outcome assessed post-exposure. Examples of this form are maternal exposure to environmental chemicals during pregnancy on birth outcomes and children's health and development \citep{wilson_2017_bias,chiu2023prenatal, hsu_prenatal_2023} and the effect of training and recovery activities over multiple days on athlete wellness \citep{schliep_2021_wellness}. A popular statistical method to estimate the time-varying association between exposure and outcome is a distributed lag model (DLM).

The DLM regresses a scalar outcome on the exposure measured at preceding time points \citep{schwartz_distributed_2000, gasparrini_distributed_2010}. The DLM framework is particularly useful as it allows for estimating time-resolved exposure effects and quantifies the temporal relationship between the exposure and the outcome. Because the repeated measurements of exposure are often correlated, DLMs typically include a smoothing constraint to add temporal structure to the estimated time-specific effects and to regularize the exposure effects in the presence of multicollinearity across the measurements. Constraints include polynomials, splines, and Gaussian processes \citep{zanobetti_generalized_2000, warren_spatial-temporal_2012}. Notably, \cite{mork_estimating_2023} introduced an approach for constrained DLM using regression tree structures based on the Bayesian additive regression tree (BART) framework \citep{chipman2010}. More recent methods have extended the tree structured DLM framework in several directions, including nonlinear exposure-response relationship \citep{mork_treed_2022}, models with lagged interaction among multiple exposures \citep{mork_estimating_2023}, and models with heterogeneous lag effects \citep{mork_heterogeneous_2023}.

There are several related methods and packages for DLM implementation. Table \ref{tab:pkgs} lists currently available R packages for DLMs that are generally appropriate for the type of epidemiology studies considered in this paper. The \CRANpkg{dlnm} package contains software for estimating a DLM with linear or nonlinear exposure-response relationships using splines for the smoothing constraint \citep{gasparrini_dlnm_2013}. \cite{mork_treed_2022, mork_estimating_2023} compare model performance between the spline-based and tree structured DLM implementations. The \CRANpkg{dlnm} package does not consider heterogeneity or multiple exposures. There are several DLM methods for the linear effect of a single exposure with modification by a single covariate. The \CRANpkg{dlim} package allows for modification by a single continuous factor \citep{demateis_penalized_2024} and the \CRANpkg{bdlim} package allows for modification by a single categorical factor \citep{wilson_bayesian_2017}.  However, these packages do not allow for multiple candidate modifying factors or multiple exposures. The package \CRANpkg{DiscreteDLM} extends Bayesian estimation to categorical outcomes but does not consider mixtures or heterogeneity \citep{dempsey2025bayesianvariableselectiondistributed}. Hence, the \CRANpkg{dlmtree} package fills several gaps including heterogeneous effects of multiple candidate modifiers and analysis of mixture or multivariate lagged exposures.

\begin{table}[ht]
\centering
\caption{\label{tab:pkgs} Available DLM related R packages with functionalities}
    \begin{tabular}{lcccc}
    Package & GLM & Nonlinearity & Mixture & Heterogeneity \\
    \hline
    \CRANpkg{bdlim}   & \checkmark  &               &            & \checkmark   \\
    \CRANpkg{DiscreteDLM}    & \checkmark  &     &            &              \\
    \CRANpkg{dlim}    & \checkmark  &               &            & \checkmark   \\
    \CRANpkg{dlmtree} & \checkmark  & \checkmark    & \checkmark & \checkmark   \\
    \CRANpkg{dlnm}    & \checkmark  & \checkmark    &            &              \\
    \hline
    \multicolumn{5}{l}{}Note: This table excludes packages designed for autoregressive DLMs,\\
    \multicolumn{5}{l}{}which are used in a different context from the models proposed here. 
    \end{tabular}
\end{table}

In this article, we introduce a comprehensive R package \CRANpkg{dlmtree} which consolidates a wide range of tree structured DLMs. The package offers a user-friendly and computationally efficient environment to fit tree structured DLMs and to address multiple potential research assumptions including nonlinearity, monotonicity and prior information, multiple simultaneous exposures, and heterogeneous lag effects. We first provide a conceptual review of a regression tree as a smoothing constraint in a DLM framework and an overview of the extensions of tree structured DLMs. We present a decision tree to help users select an appropriate model for their analysis. We illustrate the model fitting process with detailed descriptions and syntax of functions for implementing models, obtaining the summary output and inferential information, and plotting the fitted models for visualization. The package is available in the comprehensive R archive network (CRAN) and the installation instructions are provided at \url{https://danielmork.github.io/dlmtree/}.

\section{Tree structured DLMs} \label{sec:treeDLM}
\subsection{DLM tree as a smoothing constraint}

In this section, we review the regression tree approach to constrained DLM estimation, a key idea underlying the tree structured DLMs in the \CRANpkg{dlmtree} package. Classically, the DLM model is applied to time-series studies, and time is defined in terms of the outcome time. Let $y_{t}$ be the observed outcome and $x_{t}$ the observed exposure at time $t$ for $t=1,\ldots,T$. For clarity, we assume a continuous outcome when presenting the models. We discuss extensions to generalized linear models in Section \ref{sec:glm}. The DLM applied to time-series is
\begin{equation}
    \label{eqn:DLMts}
    y_{t}= \sum_{l=1}^L x_{t-l}\beta_l + \bm{z}_{t}' \bm{\gamma} + \varepsilon_{t},
\end{equation}
where $\beta_l$ is the linear effect of exposure at lag $l$ ($l$ days prior to the outcome assessment), $\bm{z}_{t}$ is a vector of covariates including the intercept, $\bm{\gamma}$ is a vector of regression coefficients for the covariates, $\varepsilon_{t}$ is independent error assumed to follow $\text{Normal}(0, \sigma^2)$.

In this paper, we focus on an alternative and more general representation of the DLM that does not assume repeated assessments of an outcome in a time-series design. Consider a vector of outcomes $\bm{y} = (y_1, \ldots, y_n)$ for a sample $i = 1, \ldots, n$. Suppose we are interested in the lagged association between the outcome and a single longitudinally assessed exposure $\bm{x}_i = (x_{i1}, \ldots, x_{iT})$ measured at equally spaced time points, $t = 1, \ldots, T$. Here, exposure time is not explicitly defined relative to the outcome. Most commonly, the exposures are in the $T$ time points prior to outcome assessment, as in \eqref{eqn:DLMts}, but that is not required. In the illustrations in this manuscript, we consider exposure during the first $T$ weeks of pregnancy and a birth or early childhood health outcome as our motivating example. Using this representation, the DLM is 
\begin{equation}
    \label{eqn:DLM}
    y_i = f(\bm{x}_i) + \bm{z}_i' \bm{\gamma} + \varepsilon_i, \quad f(\bm{x}_i) = \sum_{t=1}^T x_{it}\theta_t,
\end{equation}
where $\bm{z}_i$, $\bm{\gamma}$ and $\varepsilon_i$ are as defined above, and $f$ is a distributed lag function parameterized by $\theta_t$ representing the linear effect of exposure at time $t$. The time-series design can also be modeled using this format, and we discuss the data processing needed to convert time-series data to this format in Section~\ref{sec:dataprep}.

\begin{figure}[ht]
    \centering
    \includegraphics[width = 50mm]{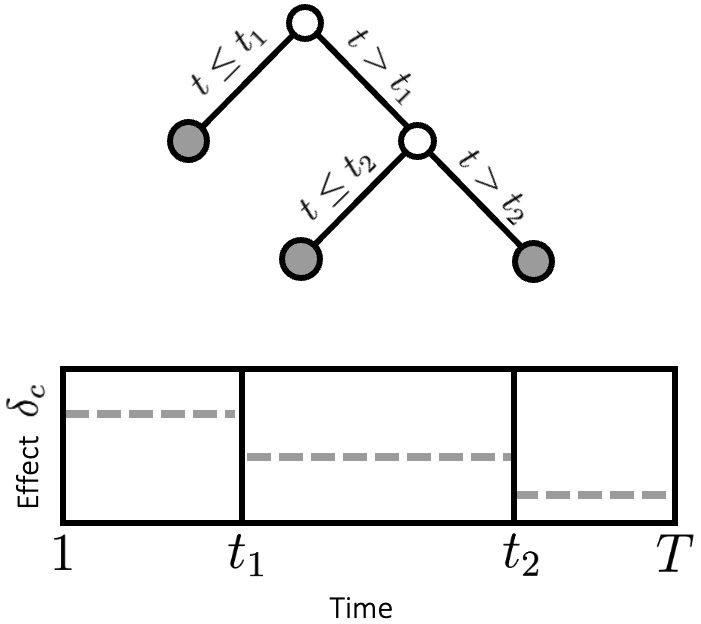}
        \caption{A DLM tree, $\tree$. A binary tree splits the time span into non-overlapping intervals, here resulting in three terminal nodes representing three time segments (gray nodes). Each terminal node is assigned a constant effect (gray dashed lines).}
    \label{fig:dlmtree}
\end{figure}

The estimation of $\theta_t$ for $t = 1, \ldots, T$ in \eqref{eqn:DLM} requires an appropriate temporal structure, such as a smooth or piecewise-smooth constraint on $\theta_t$, to account for autocorrelation within the exposure measurements. \cite{mork_estimating_2023} introduced a regression tree-structure, shown in Figure \ref{fig:dlmtree}, referred to as a \dfn{DLM tree}. 
A DLM tree, denoted $\tree$, is a binary tree that splits the exposure time span into non-overlapping segments. We denote the terminal nodes of $\tree$ as $\lambda_c$ for $c = 1, \ldots, C$ where $C$ is the total number of terminal nodes. Each terminal node of the DLM tree is assigned a scalar parameter $\delta_c$ that represents the effect of the time lags contained in that terminal node. We denote a set of scalar parameters of the DLM tree $\tree$ as $\mathcal{D} = \{\delta_1, \ldots, \delta_C\}$. A DLM tree defines a function of a time lag,
\begin{equation}
    g(t | \tree, \mathcal{D}) = \delta_c \quad \text{ if } t \in \lambda_c.
    \label{eqn:g1}
\end{equation}

The treed distributed lag model (TDLM) is a tree structured DLM in its simplest form and is a Bayesian additive model that consists of an ensemble of $A$ DLM trees, indexed as $\tree_a$, with a corresponding set of scalar parameters $\mathcal{D}_{a}$ for $a = 1, \dots, A$. TDLM defines $\theta_t$ in (\ref{eqn:DLM}) as
\begin{equation}
    \label{eqn:TDLM}
    \theta_{t} = \sum^A_{a = 1} g(t|\tree_{a}, \mathcal{D}_{a}).
\end{equation}
The representation of the DLM tree ensemble provides several advantages. Each DLM tree provides a temporal structure on the exposure-time-response function with data-driven learning of the change points and time spans related to the outcome. The ensemble structure allows for flexibility to approximate smoothness in the exposure-time-response function as each DLM tree in the ensemble splits the time span differently.

\subsection{DLM tree pair for lagged multivariate exposures} \label{sec:treepair}
A distributed lag mixture model extends the DLM framework to a multivariate exposure (also referred to as \dfn{mixture exposures} in the environmental literature) assessed longitudinally. For mixture exposures with $M \geq 2$ exposures, we denote the vector of longitudinally assessed measurements of the $m^{\text{th}}$ exposure, for $m = 1, \dots, M$, as $\bm{x}_{im} =(x_{im1}, \dots, x_{imT})$. The exposure-time-response function $f(\bm{x}_i)$ in \eqref{eqn:DLM} is replaced with a mixture-exposure-time-response function that incorporates the main effect of each exposure and the pairwise lagged interaction effects between exposures. The mixture-exposure-time-response function is
\begin{equation}
\label{eqn:DLMM}
    f(\bm{x}_{i1}, \ldots, \bm{x}_{iM}) = \sum^M_{m = 1} \sum^T_{t = 1} x_{imt}\theta_{mt} +\sum^{M}_{m_1 = 1}\sum^M_{m_2 = m_1} \sum^T_{t_1 = 1}\sum^T_{t_2 = 1} x_{im_1t_1}x_{im_2t_2}\theta_{m_1m_2t_1t_2},
\end{equation}
where $\theta_{mt}$ is a main effect of exposure $m$ at time $t$ and $\theta_{m_1m_2t_1t_2}$ is an interaction effect of exposure $m_1$ at time $t_1$ and exposure $m_2$ at time $t_2$. 

\begin{figure}[ht]
    \centering
    \includegraphics[width = 0.9\textwidth]{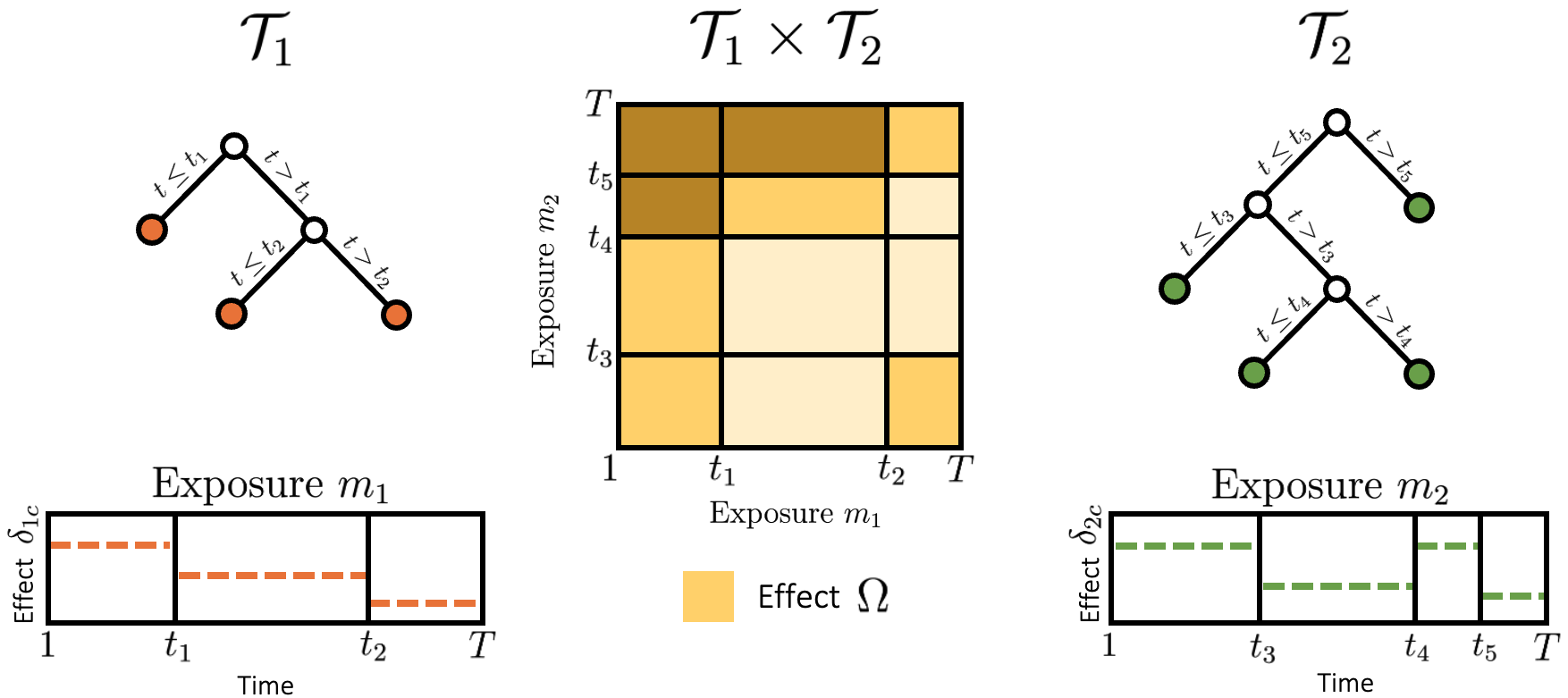}
        \caption{A DLM tree pair. Two DLM trees split the time span of the assigned exposure into non-overlapping intervals, here resulting in three time segments for exposure $m_1$ and four time segments for exposure $m_2$ (colored nodes). Each terminal node is assigned a scalar parameter that represents a constant effect of the assigned exposure (colored dashed lines). The interaction surface is fully defined by two DLM trees and each combination of time segments is assigned a scalar parameter (color-shaded boxes).}
    \label{fig:dlmtreepair}
\end{figure}

Estimating the function in \eqref{eqn:DLMM} is challenging due to autocorrelation across repeated exposure measurements, correlation at the same time lag between exposures, and a high-dimensional parameter space. \cite{mork_estimating_2023} introduced the treed distributed lag mixture model (TDLMM) that structures the parameters in \eqref{eqn:DLMM} using a \dfn{DLM tree pair}. Figure \ref{fig:dlmtreepair} illustrates a single DLM tree pair, denoted $\{\mathcal{T}_1, \mathcal{T}_2, \tree_1 \times \tree_2\}$ with corresponding parameters sets $\{\mathcal{D}_1, \mathcal{D}_2, \Omega\}$, where $\mathcal{D}_1$ and $\mathcal{D}_2$ represent the scalar main effects for $\mathcal{T}_1$ and $\mathcal{T}_2$, respectively, and $\Omega$ contains the scalar interaction effects for interaction surface, denoted $\tree_1 \times \tree_2$.
Each DLM tree in the pair is associated with one component of the mixture exposures. Similar to the DLM tree described above, each DLM tree in the pair partitions the time span of its assigned component. Each DLM tree in a tree pair is defined similarly to (\ref{eqn:g1}) such that
\begin{equation}
    \label{eqn:g2}
    g(t|\tree_{p}, \mathcal{D}_{p}) = \delta_{pc} \quad \text{ if } t \in \lambda_{pc}, \quad p = 1, 2.
\end{equation}

The DLM tree pair structure allows for an interaction surface to model lagged interactions between exposures. The interaction surface $\tree_1 \times \tree_2$ shown in the middle of Figure \ref{fig:dlmtreepair}, is fully defined by the two DLM trees in a pair. Each time interval of the first DLM tree is paired with every interval of the second DLM tree. Each combination is assigned a scalar parameter $\omega_{c_1c_2}$ that represents the lagged interaction effect where $c_1$ and $c_2$ are indices for terminal nodes of the first and second DLM tree, respectively. The interaction surface of a DLM tree pair defines a function
\begin{equation}
\label{eqn:g_I}
    g_I(t_1, t_2| \tree_{1} \times \tree_{2}, \Omega) = \omega_{c_1c_2} \quad \text{if } t_1 \in \lambda_{c_1}, t_2 \in \lambda_{c_2}.
\end{equation}
A DLM tree pair with an interaction surface provides a structure to regularize the exposure-time-response function for two components and their lagged interaction effects.

TDLMM employs an ensemble representation of $A$ DLM tree pairs denoted $\{\tree_{a1}, \tree_{a2}, \tree_{a1} \times \tree_{a2}\}$ with the corresponding set of scalar parameters $\{\mathcal{D}_{a1}, \mathcal{D}_{a2}, \Omega_a \}$, using the formulation in \eqref{eqn:g2} and \eqref{eqn:g_I}. Each DLM tree $\tree_{ap}$ is also associated with exposure $m$, indicated by $S_{ap} = m$. With TDLMM, the main effect of exposure $m$ at time $t$ in \eqref{eqn:DLMM} is 
\begin{equation}
    \label{eqn:TDLMMmain}
    \theta_{mt} = \sum^A_{a = 1} \sum^2_{p=1} g(t| \tree_{ap}, \mathcal{D}_{ap}) \mathbb{I}(S_{ap} = m).
\end{equation}
The pairwise interaction effect between exposure $m_1$ at time $t_1$ and exposure $m_2$ at time $t_2$ in \eqref{eqn:DLMM} is
\begin{equation}
    \label{eqn:TDLMMint}
    \theta_{m_1m_2t_1t_2} = \sum^A_{a=1}g_I(t_1, t_2| \tree_{a1} \times \tree_{a2}, \Omega_a) \mathbb{I}(S_{a1} = m_1, S_{a2} = m_2).
\end{equation}
TDLMM has three representations depending on different assumptions of lagged interactions between exposures. The simplest form, TDLMMadd, assumes $\theta_{m_1m_2t_1t_2} = 0$ in \eqref{eqn:DLMM}, implying no interaction between exposures resulting in an additive model of the main effects of exposures. The second form, TDLMMns, accounts for lagged interaction between exposures but not within exposures. The last, TDLMMall, allows for all lagged interactions between and within exposures, implying a nonlinear effect of exposure.

\subsection{Extensions to nonlinear exposure-time-response functions} \label{sec:nonlinear}
\cite{mork_treed_2022} introduced the treed distributed lag nonlinear model (TDLNM) to estimate the nonlinear association between a single longitudinally assessed exposure and an outcome. The DLM tree, as shown in \eqref{eqn:g1}, assumes a linear association between exposure at each time point and the outcome. To relax the linearity assumption for a distributed lag nonlinear model framework, TDLNM modifies the DLM tree to additionally split exposure concentration levels along with the exposure time span, partitioning the bi-dimensional space of exposure concentration and time lags. \cite{mork_monotone_2024} further extended TDLNM to include a monotonicity assumption, where the exposure-response is constrained to be non-decreasing at each time point. See \cite{mork_treed_2022, mork_monotone_2024} for details.

\subsection{Extensions to generalized linear models} \label{sec:glm}
In various applications of DLMs, the response variables may be binary or counts. Examples of binary outcomes include the occurrence of conditions such as asthma or preterm birth and an example of a count-valued outcome is the daily number of deaths in a county. The linear representation of the DLM framework allows the tree structured DLMs to extend to the generalized linear model setting. TDLM, TDLNM, and TDLMM have been extended to incorporate binary response variables via logistic regression \citep{mork_treed_2022, mork_estimating_2023}. Further extensions on these models allow for count data via negative binomial regression, including an option for zero-inflated negative binomial data. The extensions to binary and count data rely on a framework based on the P\'olya-Gamma data augmentation approach \citep{polson_bayesian_2013, neelon_bayesian_2019}.

\subsection{Extensions to heterogeneous models} \label{sec:het}
Another extension to the DLM framework is to assume heterogeneous exposure effects. The exposure effects may be heterogeneous due to a single modifying factor or a set of factors. For example, the impact of prenatal exposure to air pollution may be governed by genetic factors such as fetal sex \citep{rosa_association_2019}. The set of factors, referred to as \dfn{modifiers}, may be continuous, categorical, or ordinal. The general approach is to introduce an additional tree, known as a \dfn{modifier tree}, that partitions the modifier space and has a DLM tree or a DLM tree pair affixed to each terminal node. 
\begin{figure}[ht]
    \centering
    \includegraphics[width = 100mm]{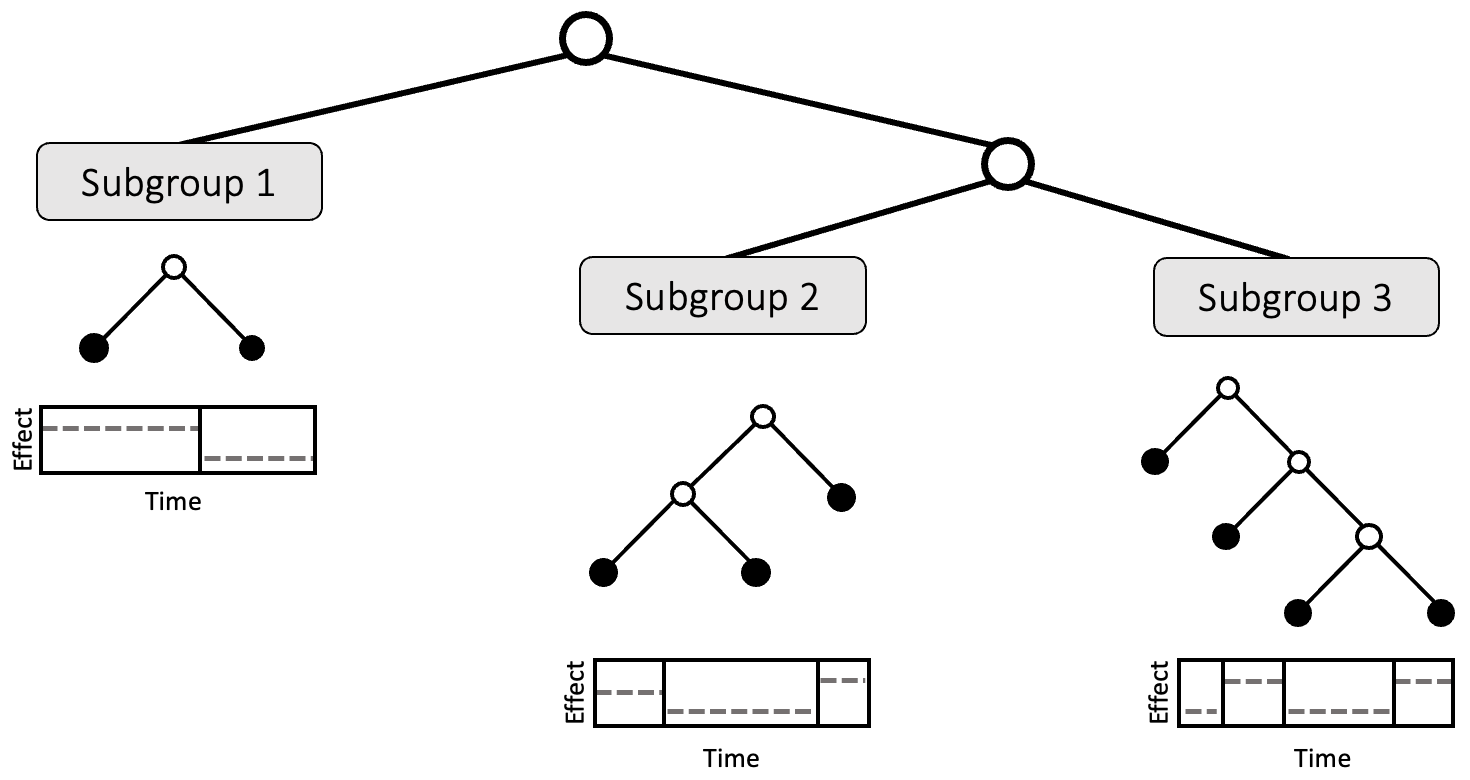}
        \caption{A nested tree structure for heterogeneous tree structured DLMs. The top tree is a modifier tree that is applied to candidate modifiers, here resulting in three subgroups. DLM trees are affixed to the terminal nodes of a modifier tree to estimate the exposure-time-response relationship specific to the subgroups.}
    \label{fig:nesttree}
\end{figure}
\cite{mork_heterogeneous_2023} extended TDLM to the heterogeneous distributed lag model (HDLM) by introducing a nested tree structure, as shown in Figure \ref{fig:nesttree}. In a nested tree, a modifier tree is applied to a set of candidate modifiers, defining mutually exclusive subgroups of the sample at each terminal node. A DLM tree is then attached to each terminal node of the modifier tree to define subgroup-specific effects with unique parameters for each subgroup. Other extensions include a heterogeneous distributed lag mixture model (HDLMM) that extends TDLMM to incorporate heterogeneity based on a set of multiple candidate modifiers using an ensemble of tree triplet structures.

\section{Implementation} \label{sec:imp}
\begin{figure}[ht]
    \centering
    \includegraphics[width = \textwidth]{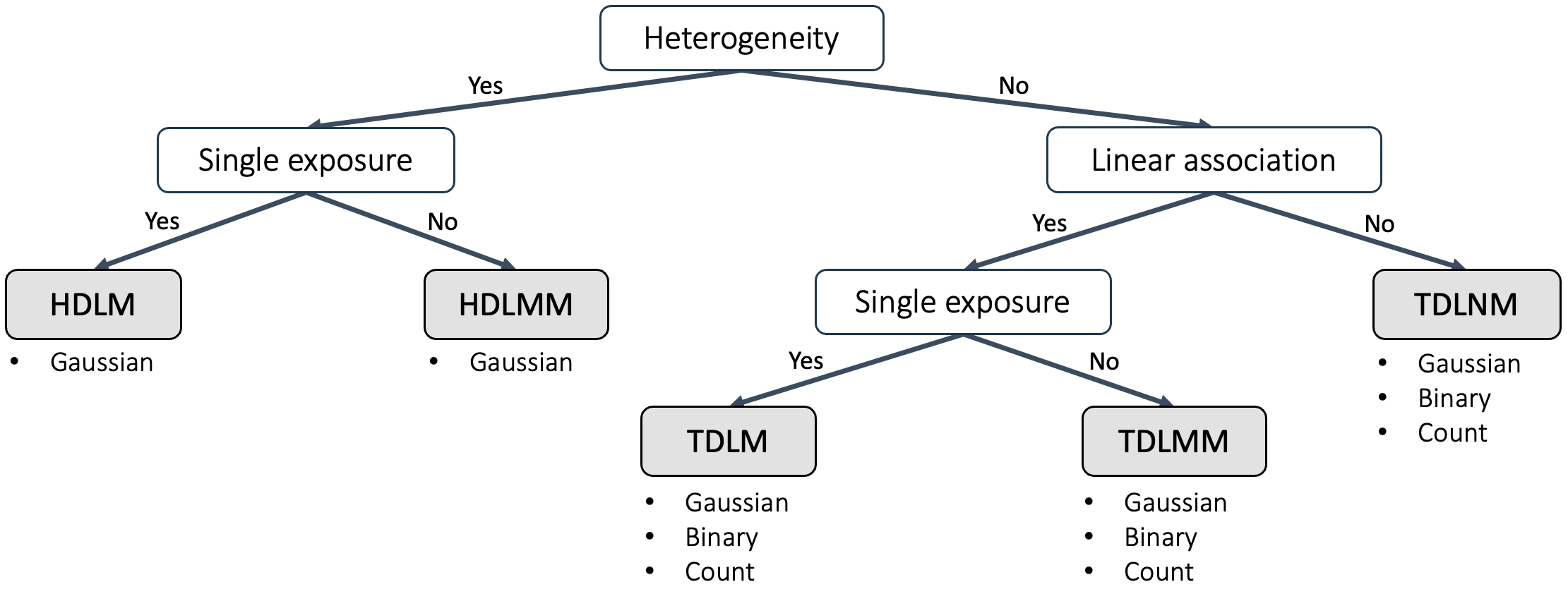}
        \caption{A decision tree for choosing tree structured DLMs. The bullet points below the models list the data types of response variables that each model can incorporate.}
    \label{fig:decisiontree}
\end{figure}

The tree structured DLMs are classified with three main criteria: 1) linear or nonlinear exposure-time-response function, 2) one lagged component or a mixture of more than one lagged component, and 3) homogeneous or heterogeneous exposure-time-response relationship. Figure \ref{fig:decisiontree} illustrates a decision tree as a guide to choosing an appropriate tree structured DLM.

The \CRANpkg{dlmtree} package offers tree structured DLMs shown in Figure \ref{fig:decisiontree}. A main function \code{dlmtree} is designed to fit various tree structured DLMs, offering customizable analysis through its arguments. The function \code{dlmtree} with its main arguments is as follows.

\begin{example}
dlmtree(formula, data, exposure.data, family, dlm.type, mixture, het)
\end{example}
The function requires a formula specifying an outcome and covariates for the fixed effect, a data source, the data type of response variable, and three arguments for model specification. The data for the \code{formula}, including covariates and the outcome, are provided as an ($n\times p$) data frame in the \code{data} argument. The exposures are not specified in the \code{formula} and are not needed in the \code{data} argument. Rather, the exposure data is specified separately as an $(n \times T)$ matrix of exposure measurements or a list of $(n \times T)$ matrices of exposure measurements in the \code{exposure.data} argument.

\begin{table}[ht]
\centering
\caption{\label{tab:dlmtreefit} Arguments for \code{dlmtree} function}
    \begin{tabular}{lllcc}
    \toprule
    Argument & Description \\
    \midrule
    \code{formula}             & Object of class \code{formula} for the fixed effect \\
    \code{data}                & A data frame containing covariates and an outcome used in \code{formula}  \\
    \code{exposure.data}       & A numerical matrix of exposure data with the same length as data. \\
                               & For a mixture setting, a named list containing equally sized numerical \\
                               & matrices of exposure data having the same length as the data  \\   
    \code{family}              & \samp{gaussian} for a continuous response, \samp{logit} for binomial, \samp{zinb} for \\
                                 & count data \\
    \code{dlm.type}            & DLM type specification: \samp{linear}, \samp{nonlinear}, \samp{monotone} \\
    \code{mixture}             & logical; A flag for mixture exposures if \code{TRUE} \\
    \code{het}                 & logical; A flag for heterogeneity if \code{TRUE}\\
    \bottomrule
    \end{tabular}
\end{table}
Table \ref{tab:dlmtreefit} provides descriptions of the arguments. All tree structured DLMs allow for continuous response variables, while TDLM, TDLNM, and TDLMM additionally allow for binary and count-valued response variables. Additional parameters, omitted here, include MCMC sampling parameters, the number of trees in the ensemble, effect shrinkage, sparsity parameters for exposure and modifier selection, and model-specific hyperparameters. These parameters can be fine-tuned and passed as a list to the corresponding \code{control} helper functions, each suffixed with its purpose (e.g., \code{control.mcmc}, \code{control.hyper}).

A fitted model is assigned a class, determined by the model-specifying arguments. The classes are: \code{tdlm}, \code{tdlmm}, \code{tdlnm}, \code{hdlm}, and \code{hdlmm}. Each class uses an S3 object oriented system with \code{summary} method. The \code{summary} method returns the model information, estimates and credible intervals for the fixed effects and lag effects, and time points of a significant effect, where the 95\% credible intervals of the lag effects do not contain zero (in some applications this is referred to as \dfn{critical window}) of the exposure or mixture exposures. The \code{plot} method on the summary object further returns the visualization of the estimated main exposure effects (and interaction effects if applicable) with 95\% credible intervals. Additionally for classes of models with heterogeneity: \code{hdlm} and \code{hdlmm}, the \code{shiny} method is built for various types of statistical inference. The \CRANpkg{shiny} app interface provides tools for identifying important modifiers and their splitting points that contribute to heterogeneity. It also includes features for evaluating personalized exposure effects with a set of user-specified modifiers, and exposure effects specific to a subgroup defined by a set of modifiers of interest.

\section{Example usage}
We illustrate the example usage of tree structured DLMs through a set of vignettes based on simulated data. We demonstrate data preparation and the model fitting process for TDLM, TDLMM, HDLM, and HDLMM. An example of fitting TDLNM is provided in the supplementary material.

\subsection{Simulated dataset}
We use the simulated dataset \file{sbd\_dlmtree}, which is publicly available in the \CRANpkg{dlmtree} package GitHub repository. The dataset contains 10,000 simulated mother-child dyads with descriptions of maternal and birth information. The maternal covariates include maternal age, height, prior weight, prior body mass index (BMI), race, Hispanic designation, education attainment, smoking habits, marital status, and yearly income. Additionally, the birth information includes birth weight for gestational age z-scores (BWGAZ), gestational age, sex of a child, and estimated date of conception. The dataset contains five environmental chemicals measured at 37 weeks preceding the birth for each dyad: fine particulate matter (PM\textsubscript{2.5}), temperature, sulfur dioxide (SO\textsubscript{2}), carbon monoxide (CO), and nitrogen dioxide (NO\textsubscript{2}). Each exposure measurement is scaled by its interquartile range value. The dataset is constructed to have realistic distributions and correlations of the covariates and exposures. It contains a complex exposure-response relationship that includes interactions and heterogeneous effects.

In the following examples, we examine the distributed lag effects of maternal exposure to the environmental mixtures during the 37 weeks of gestation on BWGAZ. The results provided in the example usage are solely for demonstrative purposes of the model fitting process using the model parameters and simulated data and do not represent any actual findings.

\subsection{Data preparation for model fitting}\label{sec:dataprep}
We first load the required packages: \CRANpkg{dlmtree} and \CRANpkg{dplyr}. We used \CRANpkg{dplyr} for a better presentation of the data processing. We set the seed for reproducibility and load the external dataset \file{sbd\_dlmtree} from \CRANpkg{dlmtree} package repository using an embedded function \code{get\_sbd\_dlmtree}.

\begin{example}
# Libraries and seed
library(dlmtree)
library(dplyr)
set.seed(1)

# Download data as 'sbd'
sbd <- get_sbd_dlmtree()
\end{example}
The data frame \code{sbd} has a lagged format where each row has columns of lagged measurements of each exposure (e.g. wide format data). Often, datasets have a time-series format, which contains a single column of dates and multiple columns of corresponding measurements of response variables and exposures on that specific date only. This is particularly common in time-series studies, whereas the wide format data is more common in cohort studies. The model fitting function \code{dlmtree} is designed to use a data frame in a wide format; hence, a data frame with a time-series format must be pivoted to a wide format. An example data frame of time-series format and a function for data pivoting are provided in the supplementary materials. In addition to the wide format, the software in this package requires that all individuals and all exposures (for multi-exposure models) have the same number of exposure time points.

To prepare the birth data in \code{sbd} for model fitting, we create a data frame including the covariates and the response variable, BWGAZ. We note that the categorical columns in the dataset are of class \code{factor}. We consider five components for exposure data: PM\textsubscript{2.5}, temperature, SO\textsubscript{2}, CO, and NO\textsubscript{2}, and store them as a list of exposure matrices with a size of $(\text{10,000} \times 37)$. Each matrix is already in wide format, where rows represent observations and columns represent exposure measurements at different lags. 

\begin{example}
# Response and covariates
sbd_cov <- sbd 
                          Hispanic, MomEdu, SmkAny, Marital, Income, 
                          EstDateConcept, EstMonthConcept, EstYearConcept) 

# Exposure data
sbd_exp <- list(PM25 = sbd 
                TEMP = sbd 
                SO2 = sbd 
                CO = sbd 
                NO2 = sbd 

sbd_exp <- sbd_exp 
\end{example}
Each matrix of the exposure data list can be centered and scaled with caution of different interpretations of the resulting estimates. Specifically when using the TDLMM with lagged interaction, it is crucial to avoid centering the exposure data as it can lead to an inaccurate estimate of the marginal exposure effect when considering co-exposures.

\subsection{TDLM: Estimating linear relationship between an outcome and a single exposure} \label{ex_tdlm}
\subsubsection{Model fitting and summary}
We first assume that we are interested in the linear association between BWGAZ and weekly exposure to PM\textsubscript{2.5} during the first 37 gestational weeks. We include the following covariates to control for fixed effects: child sex, maternal age, BMI, race, Hispanic designation, smoking habits, and month of conception. We fit TDLM with the following code.

\begin{example}
tdlm.fit <- dlmtree(formula = bwgaz ~ ChildSex + MomAge + MomPriorBMI + 
                                        Race + Hispanic + SmkAny + EstMonthConcept,
                    data = sbd_cov,
                    exposure.data = sbd_exp[["PM25"]],       # A single numeric matrix
                    family = "gaussian", 
                    dlm.type = "linear", 
                    control.mcmc = list(n.burn = 2500, n.iter = 10000, n.thin = 5))
\end{example}
The resulting fitted object of class \code{tdlm} has attributes of the fitted model information and posterior samples of parameters of interest. The \code{summary} method applied to the object \code{tdlm.fit} returns a clear overview of the model fit. The summary of the model is obtained with the following code.

\begin{example}
tdlm.sum <- summary(tdlm.fit)
print(tdlm.sum)
\end{example}

\begin{example}
---
TDLM summary

Model run info:
- bwgaz ~ ChildSex + MomAge + MomPriorBMI + Race + Hispanic + SmkAny + EstMonthConcept 
- sample size: 10,000 
- family: gaussian 
- 20 trees
- 2500 burn-in iterations
- 10000 post-burn iterations
- 5 thinning factor
- exposure measured at 37 time points
- 0.95 confidence level

Fixed effect coefficients:
                       Mean  Lower  Upper
*(Intercept)          2.289  2.032  2.542
*ChildSexM           -2.105 -2.126 -2.085
MomAge                0.000 -0.001  0.002
*MomPriorBMI         -0.021 -0.022 -0.019
RaceAsianPI           0.069 -0.057  0.192
RaceBlack             0.078 -0.050  0.205
Racewhite             0.059 -0.060  0.181
*HispanicNonHispanic  0.255  0.233  0.278
*SmkAnyY             -0.403 -0.451 -0.356
EstMonthConcept2     -0.049 -0.109  0.010
*EstMonthConcept3    -0.145 -0.211 -0.077
*EstMonthConcept4    -0.230 -0.295 -0.160
*EstMonthConcept5    -0.207 -0.265 -0.147
*EstMonthConcept6    -0.205 -0.260 -0.153
EstMonthConcept7     -0.032 -0.083  0.023
*EstMonthConcept8     0.145  0.081  0.210
*EstMonthConcept9     0.393  0.326  0.460
*EstMonthConcept10    0.372  0.311  0.437
*EstMonthConcept11    0.330  0.271  0.387
*EstMonthConcept12    0.129  0.078  0.181
---
* = CI does not contain zero

DLM effect:
range = [-0.019, 0.008]
signal-to-noise = 0.021
critical windows: 11-20,36-37 
             Mean  Lower  Upper
Period 1    0.003 -0.005  0.015
Period 2    0.000 -0.006  0.010
Period 3   -0.002 -0.010  0.004
Period 4   -0.002 -0.010  0.003
Period 5   -0.001 -0.007  0.004
Period 6   -0.001 -0.006  0.005
Period 7   -0.001 -0.006  0.006
Period 8   -0.001 -0.008  0.004
Period 9   -0.002 -0.011  0.003
Period 10  -0.002 -0.012  0.006
*Period 11 -0.016 -0.024 -0.007
*Period 12 -0.017 -0.024 -0.010
*Period 13 -0.017 -0.024 -0.012
*Period 14 -0.017 -0.022 -0.010
*Period 15 -0.017 -0.022 -0.011
*Period 16 -0.017 -0.022 -0.011
*Period 17 -0.017 -0.024 -0.011
*Period 18 -0.019 -0.030 -0.013
*Period 19 -0.018 -0.027 -0.011
*Period 20 -0.015 -0.024 -0.003
Period 21  -0.007 -0.019  0.002
Period 22  -0.002 -0.010  0.006
Period 23  -0.003 -0.012  0.003
Period 24  -0.002 -0.008  0.004
Period 25   0.000 -0.005  0.006
Period 26   0.000 -0.005  0.006
Period 27  -0.001 -0.005  0.005
Period 28  -0.001 -0.006  0.004
Period 29  -0.001 -0.007  0.004
Period 30  -0.002 -0.008  0.003
Period 31  -0.002 -0.008  0.004
Period 32  -0.001 -0.008  0.005
Period 33   0.002 -0.004  0.010
Period 34   0.004 -0.003  0.012
Period 35   0.006 -0.001  0.014
*Period 36  0.008  0.000  0.017
*Period 37  0.008  0.000  0.019
---
* = CI does not contain zero

residual standard errors: 0.004
---
\end{example}

The summary output first presents a section \samp{Model run info} with the model fitting information including the formula, sample size, data type of the response variable, number of trees in the ensemble, MCMC parameters, number of lags, and a confidence level. The next section, \samp{Fixed effect coefficients}, shows the estimates with credible intervals for the regression coefficients of the covariates. Lastly, the summary output returns \samp{DLM effect} section including the range of DLM effects, signal-to-noise ratio, and estimated lagged effects with credible intervals. Each lag is marked with an asterisk if it is identified as a critical window based on a pointwise 0.95 probability credible interval. In context, TDLM estimated a negative association between BWGAZ and PM\textsubscript{2.5} with gestational weeks 11--20 as critical windows.

The cumulative effect of the exposure is defined as the effect of a one unit increment of the exposure across all time points. The following code returns the estimated cumulative effect with its 95\% credible interval from an attribute \code{cumulative.effect} of the summary object \code{tdlm.sum}.
\begin{example}
tdlm.sum$cumulative.effect
      mean       2.5
-0.1738753 -0.2124918 -0.1367665
\end{example}
In this context, TDLM estimates that a unit increment of exposure to PM\textsubscript{2.5} across all gestational weeks has a cumulative exposure effect of -0.17 (95\% CrI: [-0.21, -0.14]) on BWGAZ.

\subsubsection{Visualizing exposure effects}
For a more intuitive view of the overall trend of the distributed lag effects, the \code{plot} method may be applied to the summary object \code{tdlm.sum} to visualize the effect estimates as the following.
\begin{example}
plot(tdlm.sum, main = "Estimated effect of PM2.5", xlab = "Time", ylab = "Effect")
\end{example}
\begin{figure}[ht]
    \centering
    \includegraphics[width=130mm]{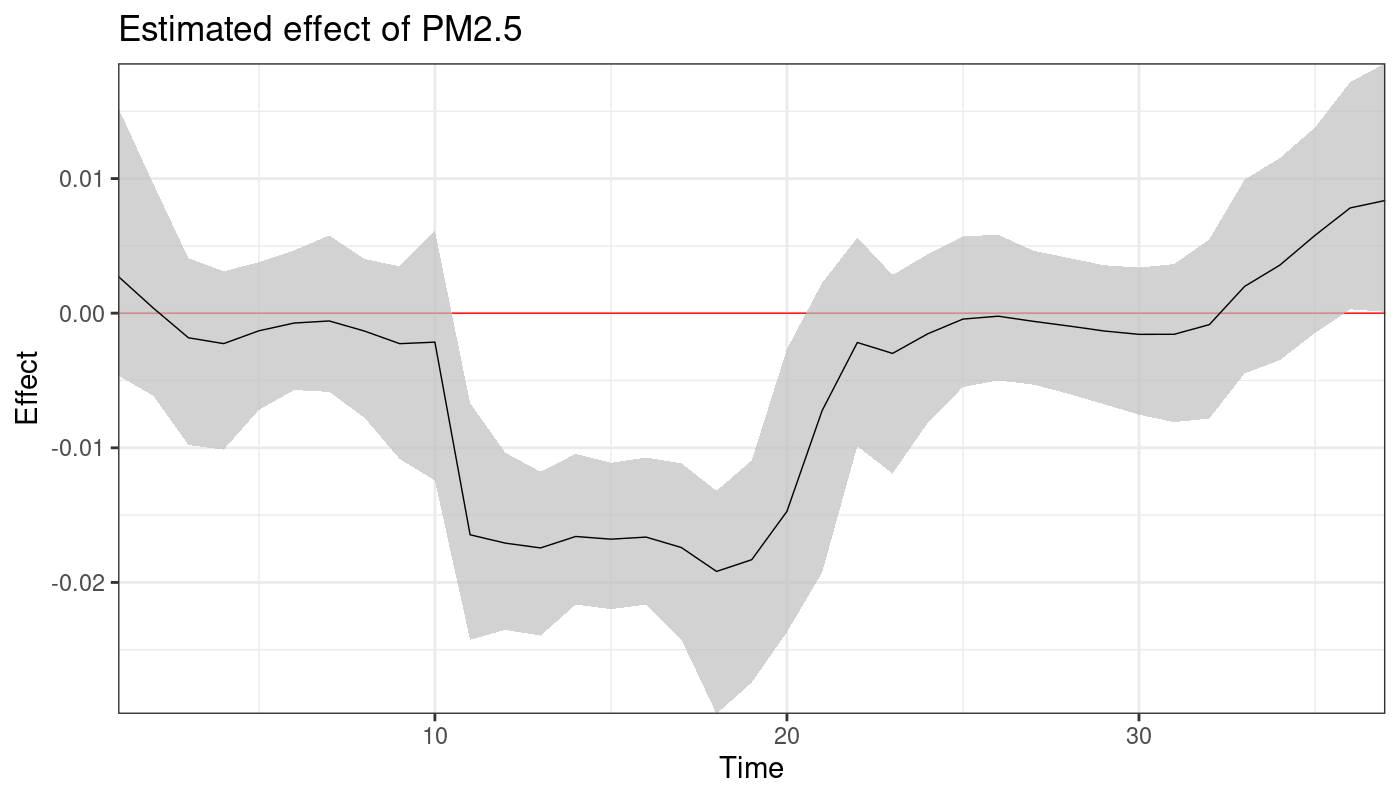}
    \caption{Estimated distributed lag effects of PM\textsubscript{2.5} on BWGAZ during 37 gestational weeks, using TDLM. These results are based on simulated data.} 
    \label{fig:tdlm}
\end{figure}
Figure \ref{fig:tdlm} shows the plot of the estimated exposure effect of PM\textsubscript{2.5} in the simulated dataset where the x-axis is the lags (or weeks) and the y-axis is the estimated exposure effect. The gray area showing the 95\% credible intervals of exposure effects during weeks 11--20 does not cover the red line of a null effect, which indicates a critical window. The \code{plot} method also includes additional arguments of \code{main}, \code{xlab}, and \code{ylab} for customizing the main title, x-axis, and y-axis label. For a customized plot, the estimated values can be obtained from the attributes of the summary object \code{tdlm.sum}: \code{matfit}, \code{cilower}, and \code{ciupper}.

\subsection{TDLMM: Analyzing linear relationship between an outcome and multiple exposures}\label{ex_tdlmm}
\subsubsection{Model fitting with pairwise lagged interactions}
Suppose we are interested in the linear association between BWGAZ and mixture exposures of five exposures: PM\textsubscript{2.5}, temperature, SO\textsubscript{2}, CO, and NO\textsubscript{2}. We are mainly interested in the marginal effects of each exposure on BWGAZ, the pairwise lagged interaction between exposures, and which exposures are most correlated with BWGAZ.  Here, we use all five exposures to illustrate the flexibility and capability of our package to handle multiple exposures. In practice, users should carefully determine the number of exposures included in the model by considering the complexity of the model relative to the number of observations and the expected signal-to-noise ratio. The key differences between the code for the model for multiple exposures below and the code for the previous single exposure model are that the \code{exposure.data} is provided with a list of matrices of five different exposures and we specify \code{mixture = TRUE}. The code is as follows.
    
\begin{example}
tdlmm.fit <- dlmtree(formula = bwgaz ~ ChildSex + MomAge + MomPriorBMI +
                                        Race + Hispanic + SmkAny + EstMonthConcept,
                     data = sbd_cov,
                     exposure.data = sbd_exp,
                     family = "gaussian",
                     dlm.type = "linear",
                     mixture = TRUE,
                     control.mix = list(interactions = "noself"),
                     control.mcmc = list(n.burn = 2500, n.iter = 10000, n.thin = 5))
\end{example}
The model assumption regarding lagged interaction can be additionally specified for the TDLMM fitting with \code{interactions} argument within \code{control.mix}. The options include no interaction (\samp{none}), no-self interactions (\samp{noself}), and all interactions (\samp{all}). We use TDLMM with no-self lagged interaction, which is the default argument. 

\subsubsection{Model summary accounting for co-exposures}
The \code{summary} method can be applied to the fitted object \code{tdlmm.fit}. As TDLMM allows for pairwise lagged interaction between mixture exposures, the estimated exposure effect for each exposure varies with the levels of the co-exposures. The \code{summary} method on class \code{tdlmm} offers additional control argument \code{marginalize} to address this. The argument \code{marginalize} requires a fixed level used for co-exposure marginalization. The default is the empirical means of co-exposures, which provides the distributed lag function for a single exposure estimated when all other exposures are fixed at their means. This is equivalent to integrating out all co-exposures. The following code returns the summary of the \code{tdlmm.fit} with marginalization using the empirical means of co-exposures.
\begin{example}
# Marginalization with co-exposure fixed at the empirical means
tdlmm.sum <- summary(tdlmm.fit, marginalize = "mean")
\end{example}
Other marginalization options are available for conducting different types of inferences. The second option is to specify a number between 0 and 100, representing a percentile of the co-exposures that will be used for marginalization. The following code returns the summary of the \code{tdlmm.fit} with marginalization fixing all co-exposures at their 25th percentile values.
\begin{example}
# Marginalization with co-exposure fixed at 25th percentile
tdlmm.sum.percentile <- summary(tdlmm.fit, marginalize = 25)
\end{example}
The last option is to specify the exact levels of co-exposures. This option requires the argument \code{marginalize} to be a numeric vector of the same length as the number of exposures used for the model fitting. The specified values in the vector must also follow the order of the exposures in the fitted model. This method of marginalization offers flexibility as these exposure levels can be specified based on pre-informed levels using existing data or to address hypothetical questions. For example, the following code can be used to obtain the marginal exposure effect of PM\textsubscript{2.5} when temperature, SO\textsubscript{2}, CO, and NO\textsubscript{2} are all fixed to 1. The marginalized effects of other exposures are calculated in a similar manner.
\begin{example}
# Marginalization with co-exposure fixed at exact levels for each exposure
tdlmm.sum.level <- summary(tdlmm.fit, marginalize = c(1, 1, 1, 1, 1))
\end{example}
Below is the summary result of \code{tdlmm.sum} with the default argument using empirical means.

\begin{example}
print(tdlmm.sum)
---
TDLMM summary

Model run info:
- bwgaz ~ ChildSex + MomAge + MomPriorBMI + Race + Hispanic + SmkAny + EstMonthConcept 
- sample size: 10,000 
- family: gaussian 
- 20 trees (alpha = 0.95, beta = 2)
- 2500 burn-in iterations
- 10000 post-burn iterations
- 5 thinning factor
- 5 exposures measured at 37 time points
- 10 two-way interactions (no-self interactions)
- 1 kappa sparsity prior
- 0.95 confidence level

Fixed effects:
                       Mean  Lower  Upper
*(Intercept)          0.172  0.043  0.307
*ChildSexM           -2.063 -2.085 -2.041
 MomAge               0.001 -0.001  0.002
*MomPriorBMI         -0.020 -0.022 -0.019
 RaceAsianPI          0.027 -0.058  0.117
 RaceBlack            0.033 -0.063  0.124
 Racewhite            0.016 -0.067  0.100
*HispanicNonHispanic  0.248  0.224  0.272
*SmkAnyY             -0.393 -0.441 -0.346
 EstMonthConcept2     0.073 -0.003  0.145
*EstMonthConcept3     0.107  0.009  0.211
*EstMonthConcept4     0.158  0.038  0.282
*EstMonthConcept5     0.255  0.126  0.388
*EstMonthConcept6     0.200  0.064  0.333
*EstMonthConcept7     0.223  0.084  0.354
*EstMonthConcept8     0.199  0.068  0.331
*EstMonthConcept9     0.291  0.164  0.418
*EstMonthConcept10    0.182  0.070  0.296
*EstMonthConcept11    0.135  0.040  0.236
 EstMonthConcept12    0.006 -0.062  0.077
---
* = CI does not contain zero

--
Exposure effects: critical windows
* = Exposure selected by Bayes Factor
(x.xx) = Relative effect size

 *PM25 (0.7): 11-20
 *TEMP (0.7): 5-19
 *SO2 (0.21): 
 *CO (0.63): 
 *NO2 (0.26): 23
--
Interaction effects: critical windows

 PM25/TEMP (0.8):
 12/6-19
 13/6-19
 14/6-20
 15/6-20
 16/6-20
 17/6-21
 18/5-22
 19/5-22
 20/6-21
---
residual standard errors: 0.005
\end{example}

The summary output of the TDLMM fit contains similar information to that of the TDLM. The \samp{Model run info} section in the output includes additional information specific to mixture exposures: the number of exposures included in the model, the number of pairwise interactions, and a sparsity parameter for exposure selection. The summary output does not include the lagged effects for each exposure to prevent overwhelming the output with excessive information. The summary presents the critical window of the marginal effects of each exposure with the relative effect size, indicating the effect size of exposure relative to that of other exposures. The summary also returns the critical window of lagged interaction effects with relative effect size. In our context, all five exposures are considered to be significantly associated with BWGAZ, based on a Bayes factor threshold of 0.5. The TDLMM estimated gestational weeks 11--20, 5--19, and 23 as critical windows of PM\textsubscript{2.5}, temperature, and NO\textsubscript{2}, respectively. The model fit also identified significant PM\textsubscript{2.5}--temperature lagged interaction effects.

\subsubsection{Additional statistical inferences using TDLMM}
More useful statistical inferences are possible with \code{tdlmm.fit} and \code{tdlmm.sum}. First, a function \code{adj\_coexposure} can be used to obtain the marginalized exposure effect while accounting for the expected change in co-exposures. In comparison to using the argument \code{marginalize} in \code{summary} method, the function \code{adj\_coexposure} uses a spline-based method to predict the expected changes in co-exposures corresponding with a pre-defined change in an exposure of interest and calculates the marginalized effects of the primary exposure with co-exposure at the predicted levels. The following code shows an example usage of the function with its output omitted.
\begin{example}
# Lower and upper exposure levels specified as 25th and 75th percentiles
tdlmm.coexp <- adj_coexposure(sbd_exp, tdlmm.fit, contrast_perc = c(0.25, 0.75))
\end{example}
The function requires exposure data, the model fit of class \code{tdlmm}, and an argument \code{contrast\_perc} which can be specified with a vector of two percentiles used for co-exposure prediction. Another argument \code{contrast\_exp} is available for specifying exact exposure levels for each exposure. 

Second, the marginal cumulative effect of each exposure can be obtained with the summary object \code{tdlmm.sum}. The object has a list attribute \code{DLM}, which contains estimates of marginal exposure effect and cumulative effect with their credible intervals. These estimated effects correspond to the argument \code{marginalize} specified within \code{summary} method. The code below returns the estimates of the cumulative effect of PM\textsubscript{2.5}.

\begin{example}
tdlmm.sum$DLM$PM25$cumulative
$mean
[1] -0.3790024

$ci.lower
      2.5%
-0.5318544 

$ci.upper
     97.5%
-0.2305391 
\end{example}
PM\textsubscript{2.5} can be replaced with other exposures if desired, e.g., \code{tdlmm.sum\$DLM\$TEMP\$cumulative} for temperature. In this context, the TDLMM estimates that the marginal cumulative effect of a unit increase of PM\textsubscript{2.5} across all gestational weeks is -0.38 (95\% CrI: [-0.53, -0.23]).

\subsubsection{Visualizing main exposure effects and lagged interaction effects}
The \code{plot} method on the summary object \code{tdlmm.sum} requires a single exposure or a pair of exposures. The \code{plot} method returns the marginal effect of an exposure when specified with a single exposure. For instance, for the three exposures, PM\textsubscript{2.5}, temperature, and NO\textsubscript{2}, we can use the following code.

\begin{example}
library(gridExtra)

p1 <- plot(tdlmm.sum, exposure1 = "PM25", main = "PM2.5")
p2 <- plot(tdlmm.sum, exposure1 = "TEMP", main = "Temperature") 
p3 <- plot(tdlmm.sum, exposure1 = "NO2", main = "NO2")

grid.arrange(p1, p2, p3, nrow = 1)
\end{example}
\begin{figure}[ht]
    \centering
    \includegraphics[width=140mm]{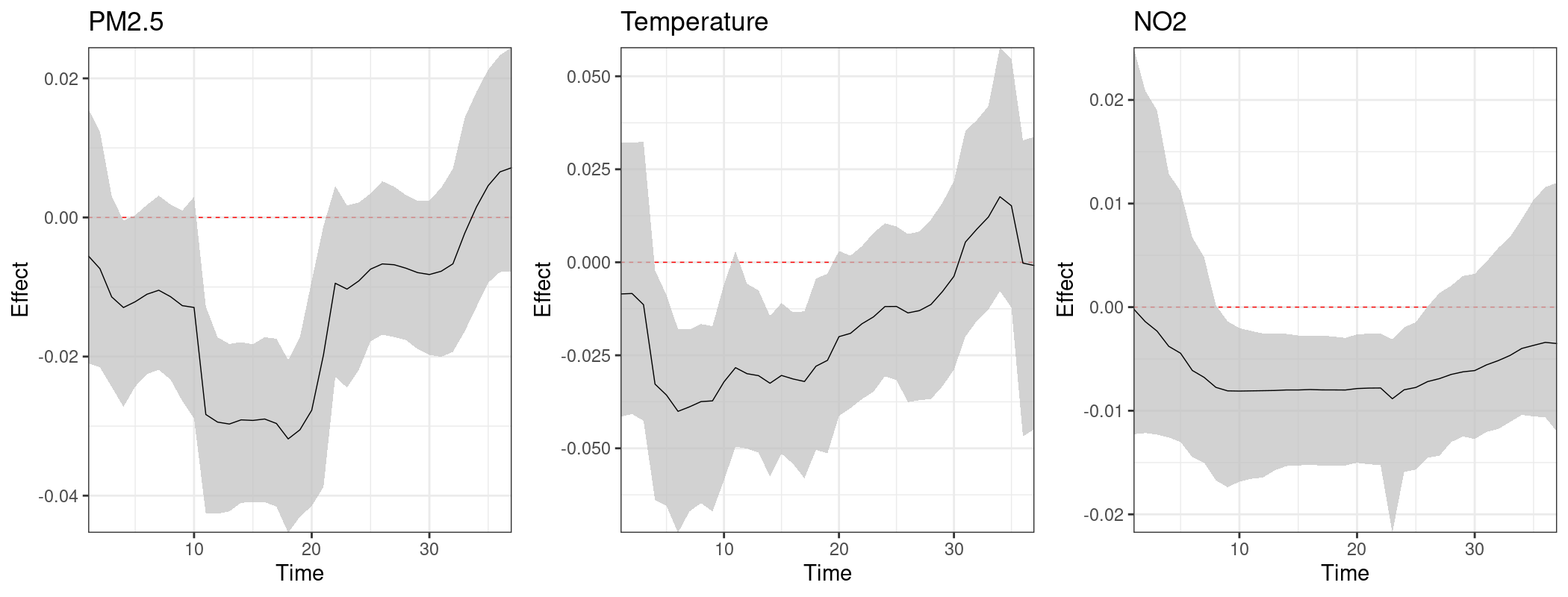}
    \caption{Estimated marginal distributed lag effects of PM\textsubscript{2.5}, temperature, and NO\textsubscript{2} on BWGAZ during 37 gestational weeks, using TDLMM. These results are based on simulated data.} 
    \label{fig:tdlmm_main}
\end{figure}
The \code{plot} method with arguments of two exposures visualizes an interaction surface of two specified exposures. The following code plots an estimated pairwise interaction surface of two specified exposures:

\begin{example}
plot(tdlmm.sum, exposure1 = "PM25", exposure2 = "TEMP")
\end{example}
\begin{figure}[ht]
    \centering
    \includegraphics[width=120mm]{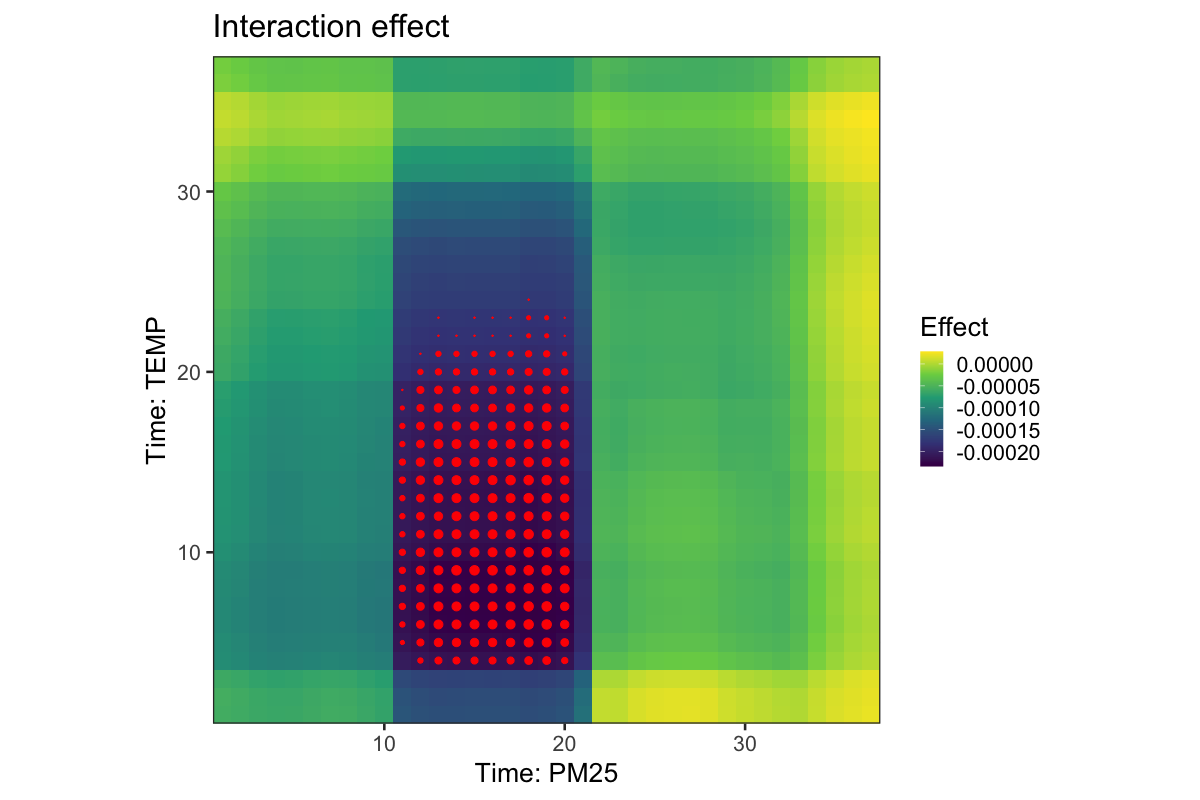}
    \caption{Estimated lagged interaction effects between PM\textsubscript{2.5} and temperature, using TDLMM. These results are based on simulated data.} 
    \label{fig:tdlmm_int}
\end{figure}
Figure \ref{fig:tdlmm_int} shows the estimated interaction surface between PM\textsubscript{2.5} and temperature in the simulated data, estimated with TDLMM. The gradient colors on the grid indicate the estimated interaction effects where the x-axis is the exposure time span of PM\textsubscript{2.5} and the y-axis is that of temperature. The red dots indicate that the credible interval of the effect does not contain zero with larger dots indicating a higher probability of non-null effect. Figure \ref{fig:tdlmm_int} indicates significant negative interaction effects at weeks 12--20 of PM\textsubscript{2.5} and 6--19 weeks of temperature at 95\% confidence level, implying that increased exposure to PM\textsubscript{2.5} may decrease the exposure effect of temperature, and vice versa.

\subsection{HDLM \& HDLMM: Introducing heterogeneity to distributed lag effects} \label{het_fit}
We illustrate heterogeneous models for a single exposure (HDLM) and for a mixture exposure (HDLMM) for estimating heterogeneous exposure effects. We focus our demonstration on a \CRANpkg{shiny} interface built for HDLM and HDLMM for examining the most significant modifying factors, the personalized exposure effects, and the subgroup-specific exposure effects. 

\subsubsection{Model fitting and summary with selected modifiers}
Suppose we are interested in the linear association between BWGAZ and a single exposure PM\textsubscript{2.5}. We additionally assume that the exposure effect of PM\textsubscript{2.5} may be modified by child sex, maternal age, BMI, and smoking habits across the population. Specifically, to fit heterogeneous models, additional arguments can be passed to \code{control.het} as a list. An argument \code{modifiers} can be set to a vector of modifier names. These modifiers must be included in the data frame provided to the \code{data} argument (\code{sbd\_cov} in this example). By default, the argument is set to include all covariates included in the \code{formula} argument as modifiers. Also, the possible number of splitting points of modifiers for heterogeneity can be specified with an integer argument \code{modifier.splits} to manage the computational cost. We specify \code{het = TRUE} to fit the HDLM with the following code.

\begin{example}
hdlm.fit <- dlmtree(formula = bwgaz ~ ChildSex + MomAge + MomPriorBMI +
                      Race + Hispanic + SmkAny + EstMonthConcept,
                    data = sbd_cov,
                    exposure.data = sbd_exp[["PM25"]],
                    family = "gaussian",
                    dlm.type = "linear",
                    het = TRUE,
                    control.het = list(
                      modifiers = c("ChildSex", "MomAge", "MomPriorBMI", "SmkAny"),
                      modifier.splits = 10),
                    control.mcmc = list(n.burn = 2500, n.iter = 10000, n.thin = 5))
\end{example}
The \code{summary} method applied to the object \code{hdlm.fit} returns the overview of the model fit. The following code returns the summary. 

\begin{example}
hdlm.sum <- summary(hdlm.fit)
print(hdlm.sum)
\end{example}

\begin{example}
---
HDLM summary

Model run info:
- bwgaz ~ ChildSex + MomAge + MomPriorBMI + Race + Hispanic + SmkAny + EstMonthConcept 
- sample size: 10,000 
- family: gaussian 
- 20 trees
- 2500 burn-in iterations
- 10000 post-burn iterations
- 5 thinning factor
- exposure measured at 37 time points
- 0.5 modifier sparsity prior
- 0.95 confidence level

Fixed effects:
                       Mean  Lower  Upper
*(Intercept)          1.272  0.914  1.629
 ChildSexM            0.127 -0.302  0.566
 MomAge               0.001 -0.002  0.004
*MomPriorBMI         -0.021 -0.024 -0.018
 RaceAsianPI          0.045 -0.074  0.171
 RaceBlack            0.055 -0.070  0.182
 Racewhite            0.034 -0.082  0.157
*HispanicNonHispanic  0.255  0.233  0.278
*SmkAnyY             -0.406 -0.453 -0.359
 EstMonthConcept2    -0.050 -0.104  0.004
*EstMonthConcept3    -0.129 -0.188 -0.070
*EstMonthConcept4    -0.201 -0.265 -0.139
*EstMonthConcept5    -0.195 -0.246 -0.144
*EstMonthConcept6    -0.199 -0.249 -0.144
 EstMonthConcept7    -0.035 -0.087  0.019
*EstMonthConcept8     0.147  0.088  0.208
*EstMonthConcept9     0.395  0.331  0.455
*EstMonthConcept10    0.388  0.328  0.446
*EstMonthConcept11    0.343  0.290  0.397
*EstMonthConcept12    0.141  0.091  0.190
---
* = CI does not contain zero

Modifiers:
               PIP
ChildSex    1.0000
MomAge      0.6305
MomPriorBMI 0.9055
SmkAny      0.0975
---
PIP = Posterior inclusion probability

residual standard errors: 0.004
---
To obtain exposure effect estimates, use the 'shiny(fit)' function.
\end{example}

As before, the summary output includes \samp{Model run info} and \samp{Fixed effects} sections with the estimates and the 95\% credible intervals of the regression coefficients of the fixed effect. The summary additionally shows the sparsity hyperparameter set for modifier selection and the posterior inclusion probability (PIP) of the modifiers included in the model. The fitted HDLM identified child sex, maternal age, and BMI to be the modifiers that contributed the most heterogeneity to the exposure effect of PM\textsubscript{2.5}. It is important to note that a high PIP does not necessarily indicate modification. Instead, it suggests that their posterior should be examined to determine if any meaningful modification is present.

A similar model fitting process can be done when examining the heterogeneous exposure effect of a mixture of five exposures on BWGAZ. The following code additionally specifies \code{mixture = TRUE} and fits HDLMM with the same potential modifiers:

\begin{example}
hdlmm.fit <- dlmtree(formula = bwgaz ~ ChildSex + MomAge + MomPriorBMI + 
                                        Race + Hispanic + SmkAny + EstMonthConcept,
                     data = sbd_cov,
                     exposure.data = sbd_exp,
                     family = "gaussian", 
                     dlm.type = "linear", 
                     mixture = TRUE, 
                     het = TRUE,
                     control.het = list(
                       modifiers = c("ChildSex", "MomAge", "MomPriorBMI", "SmkAny"),
                       modifier.splits = 10),
                     ),
                     control.mcmc = list(n.burn = 2500, n.iter = 10000, n.thin = 5))
\end{example}
As previously, the \code{summary} method on \code{hdlmm} model object similarly returns the summary of the fitted HDLMM. The summary output, omitted here, is similar to that of HDLM, with additional information such as the number of exposures, number of pairwise interactions, and sparsity parameters for modifier selection and exposure selection. Unlike the \code{summary} method applied to the model of class \code{tdlmm} in Section \ref{ex_tdlmm}, marginalization methods are unavailable for the fitted model \code{hdlmm.fit} as the marginalization of co-exposure with heterogeneity is not well defined.


\subsubsection{Using shiny app to investigate heterogeneous distributed lag effects} \label{shiny}
Exposure effects are estimated at the individual level and can be summarized at either the individual or subgroup level. This flexibility makes summarizing and visualizing the estimated effects challenging. A built-in \CRANpkg{shiny} app with an object of class \code{hdlm} and \code{hdlmm} provides a comprehensive analysis of the exposure effects. HDLM and HDLMM share the same \CRANpkg{shiny} interface but the \code{shiny} method applied to class \code{hdlmm} additionally includes an option in the panel to select an exposure of interest from mixture exposures. We present the \CRANpkg{shiny} app interface using the fitted HDLM for a single exposure, using the same argument specification for fitting the model \code{hdlmm.fit}. The \CRANpkg{shiny} app is launched with the following code.

\begin{example}
shiny(hdlm.fit)
\end{example}

\begin{figure}[ht]
    \centering
    \includegraphics[width=0.9\textwidth]{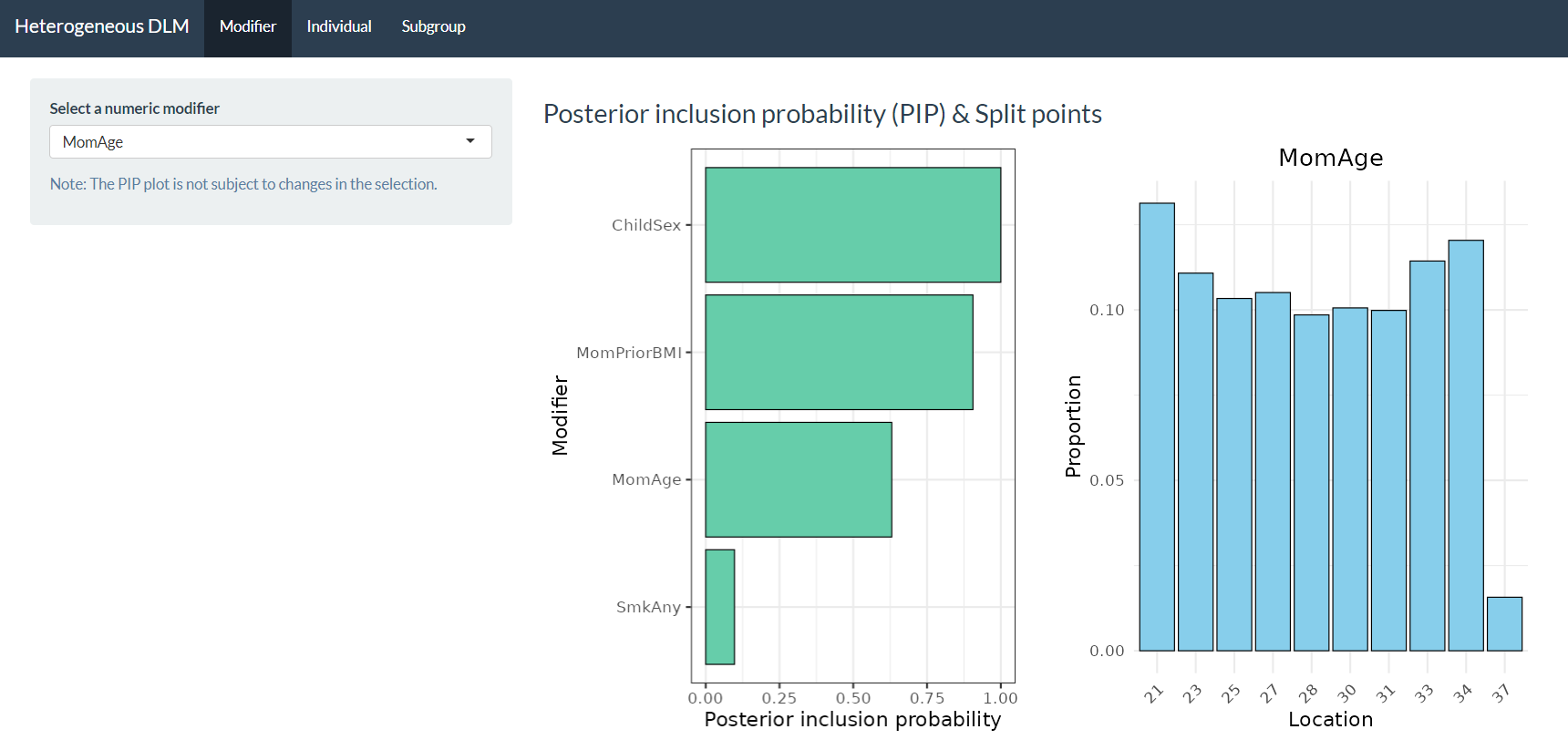}
    \caption{Example of R \CRANpkg{shiny} user interface for the fitted HDLM. The displayed results are based on simulated data.} 
    \label{fig:hdlm_shiny}
\end{figure}

Figure \ref{fig:hdlm_shiny} shows the main screen, also the first tab, of the \CRANpkg{shiny} app for the fitted HDLM. The \CRANpkg{shiny} interface includes three tabs. The first tab labeled `Modifier' presents two panels including a bar plot of modifier PIPs and the proportions of split points of a user-selected continuous modifier used to split the internal nodes of modifier trees. 

\begin{figure}[ht]
    \centering
    \includegraphics[width=0.9\textwidth]{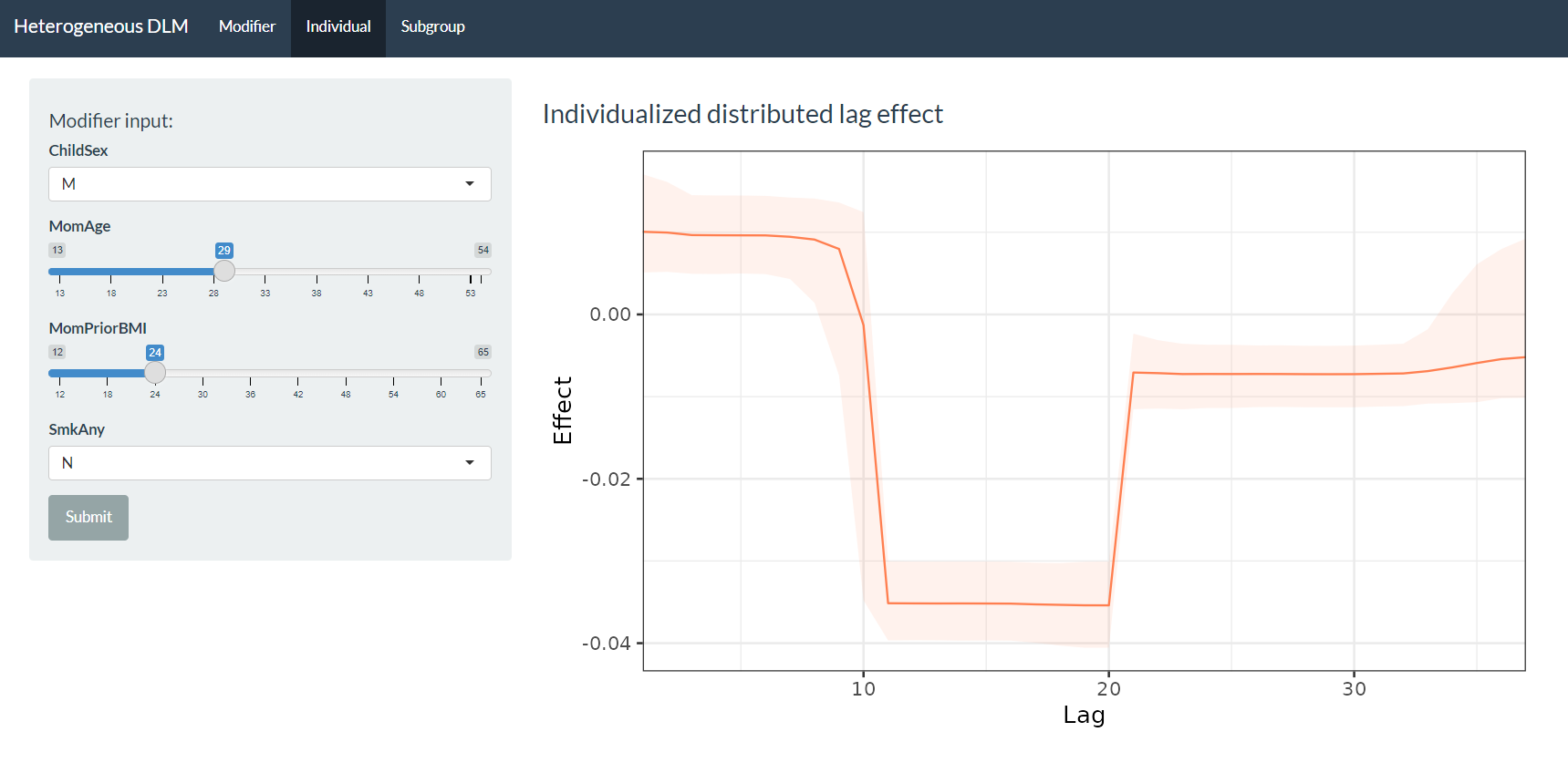}
    \caption{Example of estimated personalized exposure effect in R \CRANpkg{shiny} app. The displayed results are based on simulated data.} 
    \label{fig:hdlm_shiny2}
\end{figure}

In the `Individual' tab, the user can adjust the levels of modifiers to obtain the individualized estimate of the distributed lag effects. In our context, the \CRANpkg{shiny} app provides the personalized exposure effect and critical windows when the sex of a child, age, BMI, and smoking habits of a mother are specified. Figure \ref{fig:hdlm_shiny2} shows the estimated personalized exposure effect of PM\textsubscript{2.5} during gestational weeks for a 29-year-old mother with a BMI of 24, whose child is male and who does not smoke.

\begin{figure}[ht]
    \centering
    \includegraphics[width=0.9\textwidth]{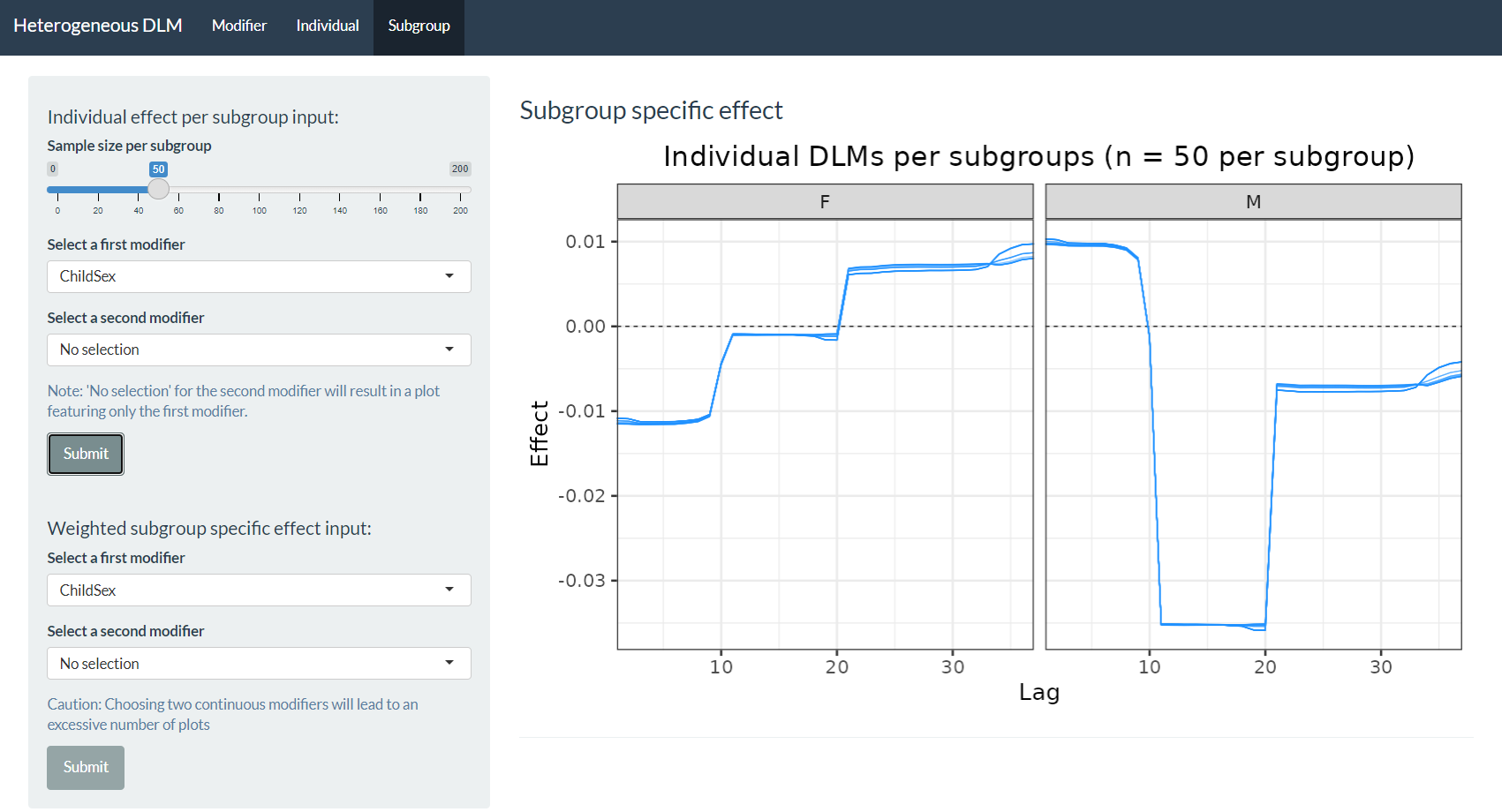}
    \caption{Example of estimated personalized exposure effects with subgroups in R \CRANpkg{shiny} app. The displayed results are based on simulated data.} 
    \label{fig:hdlm_shiny3}
\end{figure}

\begin{figure}[ht]
    \centering
    \includegraphics[width=0.82\textwidth]{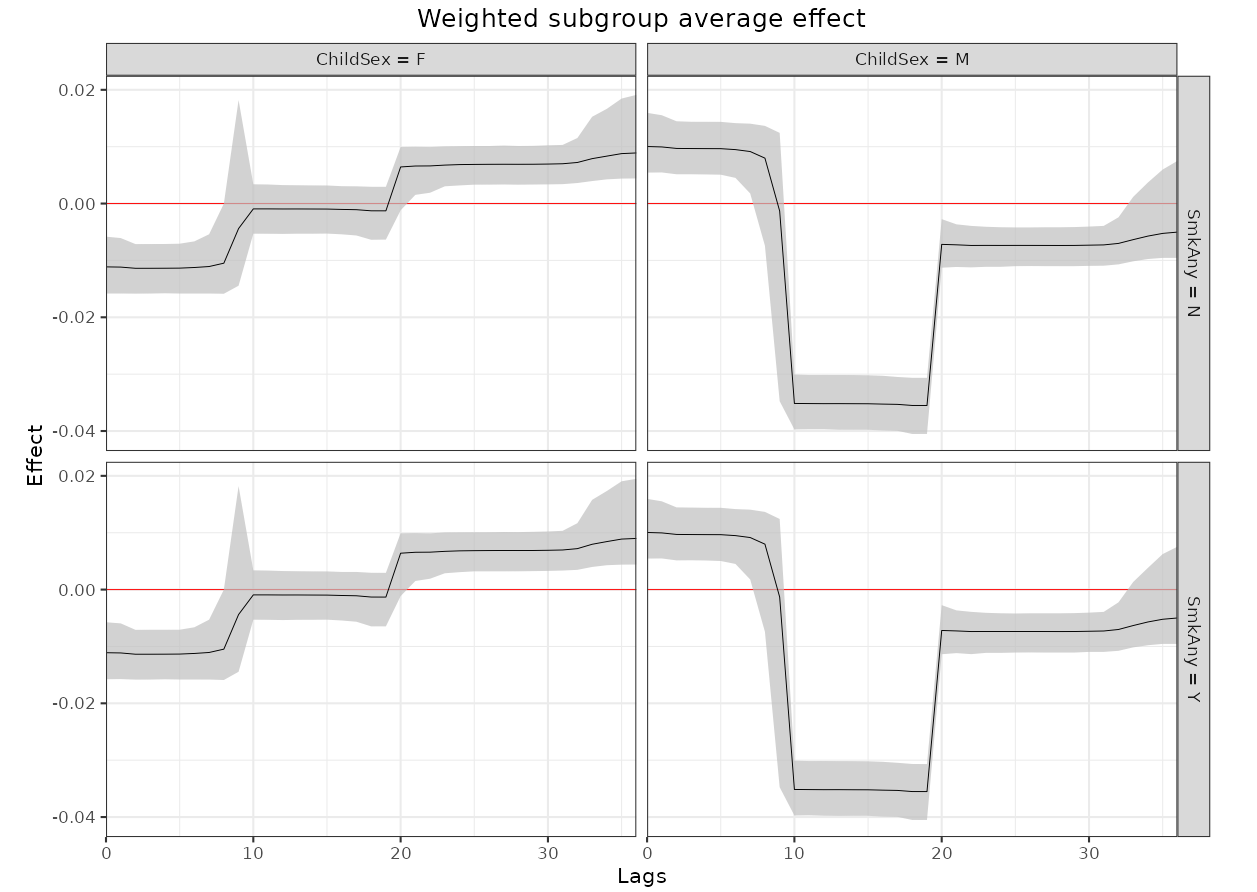}
    \caption{Example of estimated subgroup-specific effects, grouped by child sex and maternal smoking status, in R \CRANpkg{shiny} app. These results are based on simulated data.} 
    \label{fig:hdlm_shiny4}
\end{figure}

The last tab labeled `Subgroup' offers subgroup-specific analyses. In the top panel, shown in Figure \ref{fig:hdlm_shiny3}, the user can select one or two modifiers to group the samples into multiple subgroups and obtain personalized exposure effects for individuals in each subgroup. Each line of the resulting plot represents an exposure effect of an individual accounting for all modifiers of that individual. This is useful for simultaneously assessing how much exposure effects vary among individuals and across subgroups.
The bottom panel allows users to select one or two modifiers for subgroup-specific distributed lag effects. Subgroup-specific distributed lag effects are calculated by marginalizing out the modifiers not specified in the panel. Two modifiers at most can be specified for analyzing how much each modifier affects heterogeneity in the exposure effects. For example, Figure \ref{fig:hdlm_shiny4} shows the estimated exposure-time-response function for four subgroups, grouped by two categorical modifiers: child sex and smoking habit. The difference in subgroup-specific exposure effects indicates that smoking habit does not contribute much heterogeneity in the exposure effect of PM\textsubscript{2.5} while child sex introduces a considerable amount, which aligns with the modifier PIPs in the summary output. Using the simulated dataset, the subgroup-specific effects suggest that mothers with a male child may be more vulnerable to PM\textsubscript{2.5}.

\newpage

\section{Practical considerations}\label{sec:practical}
R package \CRANpkg{dlmtree} offers highly flexible and integrative models for analyzing the effects of repeatedly measured exposures. The package allows user specification for adjusting the flexibility of the exposure-lag-response relationship, number of exposures, types of lagged interaction terms, modifiers, and MCMC simulation control. We highlight important considerations and best practices when using the package to obtain reliable results.

The models included in \CRANpkg{dlmtree} are very flexible and highly parametric. It is, therefore, important to be mindful of whether such a complex model is relevant to a research question. Factors that increase model complexity are: increased number of time points the exposure is assessed at, more unique exposures and allowing for additional interactions between or within exposures across time, and including heterogeneity and the number of candidate modifiers. The BART framework and shrinkage and selection priors are effective at regularizing these models. However, more complex models can still result in increased variance, identifiability issues where the model becomes sensitive to model specification, and challenges in interpretation. Therefore, we recommend users employ standard best practices in model building such as carefully assessing what should be included in the model, considering pre-selection of exposures or combining related exposures into a single index, aggregating exposure measurements to a coarser time interval, being mindful of extreme multicollinearity, and conducting sensitivity analyses with simpler models where possible.

When running heterogeneous distributed lag models, potentially important modifiers can be identified using the PIP.  When the number of modifiers is small relative to the number of trees, all modifiers will have a higher baseline PIP. Additionally, a large PIP does not necessarily imply meaningful effect modification and the posterior distribution of the exposure-response function should be explored to better identify important modifiers. Related to model complexity, when two or more modifiers are highly correlated one or both may be identified as driving heterogeneity. Therefore, considering the mechanisms of effect modification is essential when selecting modifiers and interpreting results.

For practical implementation, we recommend running longer MCMC chains than are used in the examples in this paper. While the examples presented use short iterations for demonstration, longer chains are necessary for stable estimates and to improve mixing and convergence. Given the high dimensionality of the tree structured DLMs, a longer chain will sample more reliable full posterior distributions for inference. Additionally, for logistic and negative binomial models longer burn-in periods are important to ensure convergence. For example, in our previous large data analyses with this software \citep{mork_estimating_2023,mork_heterogeneous_2023,mork_treed_2022}, we have used upwards of 10,000 burn-in iterations. 

Convergence can be assessed in several ways. We have relied on trace plots, which graph the chain of posterior draws for key parameters. When possible, another approach is to run multiple Markov chains with different seeds and compare results across chains. To help users assess model convergence and identify potential issues with mixing, our package provides a \code{diagnose} S3 method for model summary objects which returns a comprehensive convergence diagnostics panel customized for the treed DLM framework. It provides an overview of key outputs including trace plots and posterior density plots of distributed lag effects, Metropolis-Hastings acceptance rate for tree updates, and changes in tree sizes throughout MCMC sampling. The example usage of \code{diagnose} is demonstrated in the supplementary materials.

\section{Summary}
We introduced the R package \CRANpkg{dlmtree}, a user friendly software for addressing a wide range of research questions regarding the relationship between a longitudinally assessed exposure or mixture and a scalar outcome. Our software provides functionality to estimate distributed lag linear or nonlinear models, quantify main and interaction effects, and account for heterogeneity in the exposure-time-response function. Furthermore, the methods in our package have been carefully optimized for computation efficiency under a custom C\texttt{++} language framework with a convenient R wrapper function, \code{dlmtree}, designed for accessibility by all researchers. In this paper, we provided an overview of the regression tree approaches used for estimating DLMs, and through a collection of vignettes, we highlighted a variety of tools available for processing data, fitting models, conducting inference, and visualizing the results. Our goal in making this package available is to bring robust data science tools to expand the range of questions that can be asked and answered with longitudinally assessed data.

\section{Acknowledgements}

Research reported in this publication was supported by National Institute of Environmental Health Sciences of the National Institutes of Health under award numbers ES035735, ES029943, ES028811, AG066793, and ES034021. The content is solely the responsibility of the authors and does not necessarily represent the official views of the National Institutes of Health.

\bibliography{main}

\address{Seongwon Im\\
  Department of Statistics\\
  Colorado State University\\
  United States of America\\
  (0009-0000-8447-5852)\\
  \email{seongwon.im@colostate.edu}}

\address{Ander Wilson\\
  Department of Statistics\\
  Colorado State University\\
  United States of America\\
  (0000-0003-4774-3883)\\
  \email{ander.wilson@colostate.edu}}

\address{Daniel Mork\\
  Department of Biostatistics\\
  Harvard T.H. Chan School of Public Health\\
  United States of America\\
  (0000-0002-7924-0706)\\
  \email{dmork@hsph.harvard.edu}}
  
\end{article}

\end{document}